\documentclass[aps,prl,reprint,superscriptaddress,longbibliography]{revtex4-2}
\usepackage{amsbsy}
\usepackage{amssymb}
\usepackage{amsmath}
\usepackage{graphicx}
\usepackage{dcolumn}
\usepackage{bm}
\usepackage[colorlinks,citecolor=blue,urlcolor=blue,bookmarks=false,hypertexnames=true]{hyperref} 

\begin{document}

\preprint{APS/123-QED}

\title{Magnetic Gradiometry from Parametric  Sideband Generation}
\title{Intrinsic Pulsed Magnetic Gradiometer in Earth's Field}

\author{Kaleb Campbell}
\affiliation{%
Sandia National Laboratory, 1515 Eubank SE, Albuquerque, NM 87123 }%
\affiliation{%
Center for Quantum Information and Control,  Department of Physics and Astronomy, University of New Mexico, Albuquerque, NM 87131}%

\author{Ying-Ju Wang}%
\affiliation{%
 QuSpin Inc, 331 S 104th St. Unit 130, Louisville, CO 80027 }%

\author{Igor Savukov}
\affiliation{%
  Los Alamos National Laboratory, Los Alamos, NM 87545}%

\author{Peter Schwindt}%
\affiliation{%
Sandia National Laboratory, 1515 Eubank SE, Albuquerque, NM 87123 }%

\author{Yuan-Yu Jau}
\affiliation{%
Sandia National Laboratory, 1515 Eubank SE, Albuquerque, NM 87123 }%

\author{Vishal Shah}
\affiliation{%
 QuSpin Inc, 331 S 104th St. Unit 130, Louisville, CO 80027 }%

 \homepage{http://www.Second.institution.edu/~Charlie.Author}

\date{\today}

\begin{abstract}
We describe a novel pulsed magnetic gradiometer based on the optical interference of sidebands generated using two spatially separated alkali vapor cells. The sidebands are produced with high efficiency using parametric frequency conversion of a probe beam interacting with $^{87}$Rb atoms in a coherent superposition of magnetically sensitive hyperfine ground states. Interference between the sidebands generates a low-frequency beat note whose frequency is determined by the magnetic field gradient between the two vapor cells. In contrast to traditional magnetic gradiometers, our approach provides a direct readout of the gradient field without the intermediate step of subtracting the outputs of two spacially separated magnetometers. The technique is expected to provide effective common-mode magnetic field cancellation at frequencies far greater than the bandwidth of the gradiometer. Using this technique, we developed a compact magnetic gradiometer sensor head with integrated optics with a sensitivity of $25 \ fT/cm/\sqrt{Hz}$ with a $4.4$ cm baseline, while operating in a noisy laboratory environment unshielded from Earth's field. We also outline a theoretical framework that accurately models sideband generation using a density matrix formalism. 
\end{abstract}

\maketitle

\begin{figure}[t]
\begin{center}
\includegraphics[width=.5\textwidth, height = .085\textwidth]{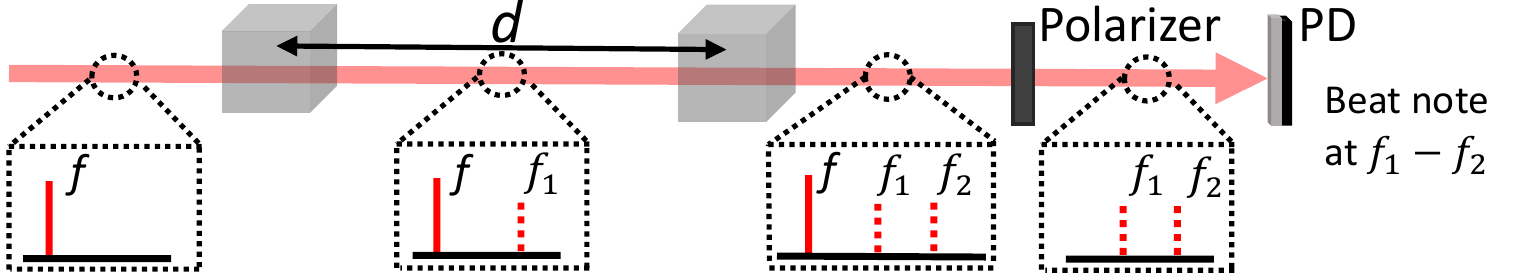}
\end{center}
\vspace{-0.25in}
\caption{\label{fig:epsart}(a) Conceptual overview of the gradiometer. Two alkali vapor cells are separated by a distance, $d$. A linear probe (carrier) beam with frequency, $f$, passes through the two vapor cells. The probe beam's interaction with the atoms in Cell $1$ produces an orthogonally polarized optical sideband at magnetic-field-dependent frequency $f_1$. Similarly, interaction with Cell $2$ produces a second optical sideband, at frequency $f_2$. The probe beam is removed using a polarizer, leaving behind only the two optical sidebands. The sidebands are captured by a photodetector (PD) where they interfere to produce a beat note at frequency $f_1-f_2$ that is directly proportional to the magnetic gradient field between the two vapor cells.}
\label{fig:concept}
\end{figure}


The last two decades have witnessed steady progress in the field of optically pumped magnetometers (OPMs) \cite{bell1957,dehmelt1957,budker2007, Tierney} based on alkali vapor cell technology. The sensitivity of OPMs \cite{kominis2003,dang2010} now rivals superconducting quantum interference device (SQUID) \cite{cohen1972} magnetic sensors that have long been the gold standard for biomagnetic measurments \cite{hamalainen,vrba}. The sensitivity of these magnetometers is far below the magnetic noise floor set by ambient geophysical magnetic activity and the urban environment. A widely used technique to remove environmental noise is using a magnetic gradiometer \cite{hamalainen} in which the output of two sensors is subtracted, leaving behind the signal of interest. With SQUIDs, two spatially separated flux pickup coils with opposing polarities can be wired together to form an intrinsic gradiometer \cite{Zimmerman,Koch}. In contrast, few OPM-based techniques exist to build similar intrinsic gradiometers using vapor cells \cite{Kamada15,sulai2019, zhang2020,perry2020all,Rom2021}. Most OPM-based gradiometers rely on subtracting the output of two spatially separated magnetometers (synthetic gradiometer) \cite{shah2013,Johnson2010,limes2020}, and differences and drifts between the two magnetometers can reduce the rejection of common-mode signals. Here, we demonstrate an unshielded intrinsic optical gradiometer where the atom-optical system itself performs the gradient subtraction between two spatially separated vapor cells, which in principle should provide good common mode rejection across a broad frequency range. 

\begin{figure}[t]
\begin{center}
\includegraphics[width=.5\textwidth, height=.33\textwidth]{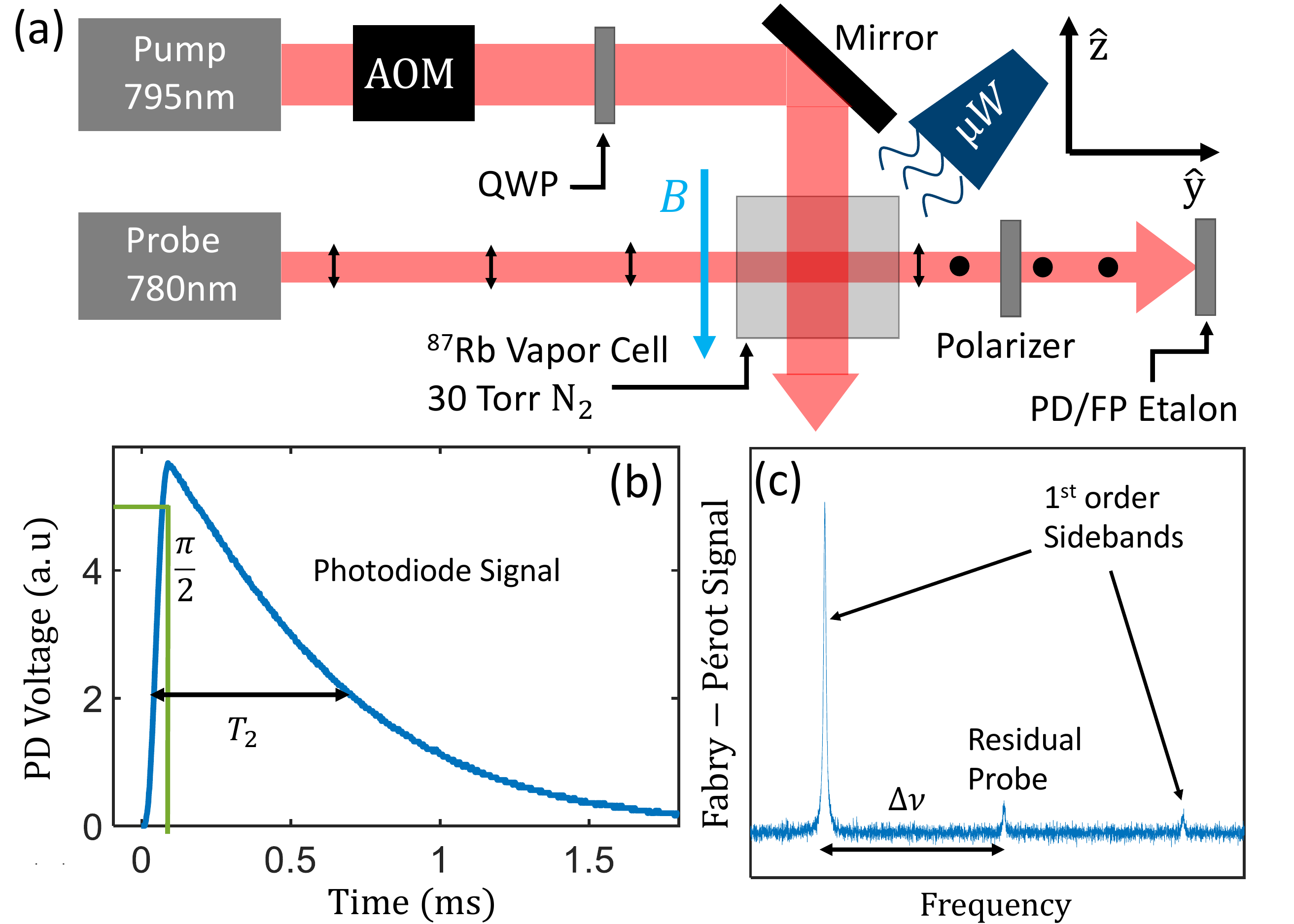}
\end{center}
\vspace{-0.25in}
\caption{\label{fig:epsart}(a) Experimental setup to observe sideband generation. QWP: quarter wave plate. $\mu W$: microwave horn. AOM: acousto-optic modulator. (b) An example of a sideband measured with the PD. The decoherence time is denoted as $T_2$. (c) PD is replaced with a Fabry-Perot etalon to verify probe/sideband extinction depending on polarizer angle.}
\label{fig:concept}
\end{figure}

Other highlights of our approach are (i) the system can operate with near-unity spin polarization, thus maximizing signal generation, and (ii) our method allows for measurement of signal photons with strong suppression of background photons which removes excess noise contribution from the background light. These two features remove fundamental barriers for reaching atom shot-noise-limited sensitivity \cite{braginsky,kominis4}. In addition, our technique probes the hyperfine magnetic end states which amplifies the sensitivity of the energy levels to magnetic fields by a factor of three compared to the $\Delta m_F \pm 1$ Zeeman states utilized in traditional alkali optically pumped magnetometers \cite{kimball2013}.
 
 The magnetic gradiometer technique is based on the generation of magnetic-field-dependent optical sidebands (Fig.~1). The sideband generation process is inspired by experiments performed in the 1970s which showed that alkali atoms prepared in a coherent superposition of hyperfine ground states can modulate a weak probe (carrier) beam in a process known as parametric frequency conversion \cite{tang1970parametric,tang1973parametric}. The time-dependent phase of the atomic coherence oscillates at the hyperfine resonance frequency, modulating the optical susceptibility of the atomic medium for near resonance light. When the atoms are probed by a weak beam, the oscillating susceptibility generates optical sideband(s). The sideband generation technique can be adopted for a number of different applications such as bio-magnetic field sensing \cite{boto2018,borna2020,jensen2018,jensen20}, timing \cite{knappe2004,camparo2007}, and microwave detection \cite{gerginov,horsley2016}.

We use the $ 5s^{2} S_{1/2} |F = 1, m_F =1\rangle = |1, 1\rangle$  and $|F = 2, m_F =2\rangle = | 2, 2\rangle$  ground state sublevels in a warm ensemble of  $^{87}$Rb as the basis for the gradiometer. Fig. 2a shows our tabletop experimental setup demonstrating sideband generation. A cubic vapor cell with internal dimensions of $8\times8\times8$ mm$^{3}$ is filled with enriched $^{87}$ Rb and $30$ Torr of nitrogen $(N_2)$ buffer gas to minimize depolarization from cell wall collisions and to limit radiation trapping \cite{rosenberry2007}, which limits efficient optical pumping \cite{RevModPhys}. The vapor cell is exposed to a background magnetic field, $\bold{B}$, along the $\hat{z}$ direction. A circularly polarized ($\sigma^{+}$) 795 nm laser tuned to the $5s^{2}S_{1/2}\rightarrow 5p^{2}P_{1/2}$ $D1$ line, also along the $\hat{z}$ direction, optically pumps over $95\%$ of the atoms into the $|2, 2\rangle$ ``dark" state. As the pumping approaches equilibrium, the pump light is switched off and the vapor cell is irradiated with a short ($30 \ \mu s$) microwave $\pi/2$ pulse using a microwave horn to prepare the atoms in a coherent supperposition of the $|1, 1\rangle \leftrightarrow  |2, 2\rangle$ hyperfine ground states. A linearly polarized $780$ nm laser (probe) tuned near the $|F = 1\rangle = |1\rangle \rightarrow  5p^{2} P_{1/2} $ D$2$ line propagates through the vapor cell in a direction orthogonal to the pump. The interaction between the probe and the coherently prepared atoms generates an optical sideband near the $|F=2\rangle =|2\rangle \rightarrow 5p^{2} P_{1/2}$ transition. Sideband generation begins immediately after the $\pi/2$ pulse is applied, and its amplitude exponentially decays on a timescale $T_2$, the ground state coherence lifetime of the rubidium atoms in the vapor cell (Fig. 2b). The generated first-order sideband(s) have linear polarization which is orthogonal with respect to the probe (carrier). The polarization orthogonality allows the probe and sidebands to be easily separated with a simple polarizer and analyzed independently. This was explicitly verified by replacing the photodiode with a scanning Fabry-Perot (FP) etalon (Fig. 2c) and  extinguishing the probe and sidebands selectively by rotating the polarizer by a $90^{\circ}$ offset angle. The first order sidebands shown in Fig. 2c were generated immediately after the $\pi/2$ pulse was applied and disappeared once the rubidium ground state coherence dissipated. 

Sideband generation is dependent on several parameters such as the degree of spin polarization, buffer gas pressure, probe detuning from resonance, and rubidium density (vapor cell temperature). Experimental observations of parameter dependency closely match predictions from our theoretical model described below. Under optimal conditions, the sideband amplitude is roughly equal to the probe amplitude after the beam exits the vapor cell. We found between $10$ and $30$ Torr of nitrogen buffer gas pressure to be ideal for sideband generation for our vapor cell dimensions. Pressure broadening \cite{PhysRevA} of the $5p^{2} P_{3/2}$ excited state due to buffer gas at pressures higher than $30$ Torr causes off-resonant excitation of the $|2\rangle$ state by the probe which reduces sideband generation efficiency, while buffer gas pressure lower than $10$ Torr increases ground state relaxation through rubidium collisions with the vapor cell walls. The temperature of the vapor cell is adjusted to obtain a rubidium optical depth (OD) $\approx 1$. In experiments at OD $>1$, the probe is detuned from $|2\rangle \rightarrow | 5p^{2} P_{3/2}\rangle$ to maximize sideband generation.

The frequency difference between the sideband and the probe, $\Delta v$, in low magnetic field is given by, 

\begin{equation}\label{v}
\Delta \nu = \Delta \nu_{HFS} + \nu_{BG} + 3  \gamma\mid{\bold{B}}\mid
\end{equation}
where $\Delta \nu_{HFS}$ is the separation between the two hyperfine ground states $|1\rangle$ and $|2\rangle$, $\nu_{BG}$ is the buffer gas induced pressure shift between the ground states, $\gamma$ is the gyromagnetic ratio  ($2\pi \cdot 6.99 \ \frac{\textrm{Hz}}{\textrm{nT}}$), and $\mid \bold{B} \mid$ is the absolute background magnetic field experienced by the vapor cell. The dependence of the difference frequency, $\Delta \nu$, on the magnetic field is amplified by a factor of three by probing the $|1,1\rangle \leftrightarrow |2,2\rangle$ hyperfine ground states.

To quantitatively analyze the experimental results, we developed a numerical model to simulate the sideband generation phenomena shown in Fig.~2. Using the previously developed theoretical framework described in an unpublished manuscript \cite{Jau2007}, or chapter 8, Ref.\cite{happer2010optically}, we find an equation for a forward-propagated electric field through an atomic medium with multi-frequency components as

\begin{equation}\label{EP}
\frac{\partial \widetilde{\mathbf{E}}({\zeta}) }{\partial {\zeta}}=-i\mathbf{K}\cdot (\mathbf{1}+\frac{\boldsymbol{\chi}}{2}) \cdot\widetilde{\mathbf{E}}({\zeta}),
\end{equation}
where ${\zeta}$ is the propagation distance through the medium in the direction of the probe, $\mathbf{K}$ is a k-vector dyadic operator (second order tensor), $\mathbf{1}$ is the unit dyadic, $\boldsymbol{\chi}$ is the electric susceptibility dyadic operator under the assumption $|\boldsymbol{\chi}|\ll1$, and $\widetilde{\mathbf{E}}({\zeta})$ is the frequency-quantized, complex position dependent electric field. A derivation of the propagation equation is given in the supplementary material, as well as an explanation of the matrix elements of the multifrequency dyadics shown in Eq.~(2) 

In the model, we use the two stretched ground-state hyperfine levels, $|1,1\rangle$ and $|2,2\rangle$. The other ground state sublevels have insignificant contributions to the sideband generation process if the optical pumping is efficient. One can think of the sideband generation process as stimulated Raman transitions between ground and excited states, forming a $\Lambda$ system with an incident photon and a stimulated photon. For the experimental configuration of a linearly polarized probe which has its propagation direction orthogonal to the magnetic field, $\sigma^{+}, \sigma^{-}$, and $\pi$ transition can occur. Fig. 3a shows a possible $\Lambda$ configuration for the negative frequency sideband, with two other configurations shown in the inset. The probe makes the transitions $|1,1 \rangle \xrightarrow{\sigma^{+}} |2',2 \rangle \xrightarrow{\pi \ } |2,2 \rangle$, $|1,1 \rangle \xrightarrow{\pi \ } |2',1\rangle \xrightarrow{\sigma^{-}} |2,2 \rangle$, and $|1,1 \rangle \xrightarrow{\pi \ } |1',1 \rangle \xrightarrow{\sigma^{-} } |2,2 \rangle$ where the prime indicates the excited state. For a 1st-order effect, a probe photon ($\sigma$ polarization) is scattered into the first order sideband ($\pi$ polarization), which is orthogonal to the probe polarization. 

\begin{figure}[t]
\begin{center}
\includegraphics[width=.5\textwidth, height=.575\textwidth]{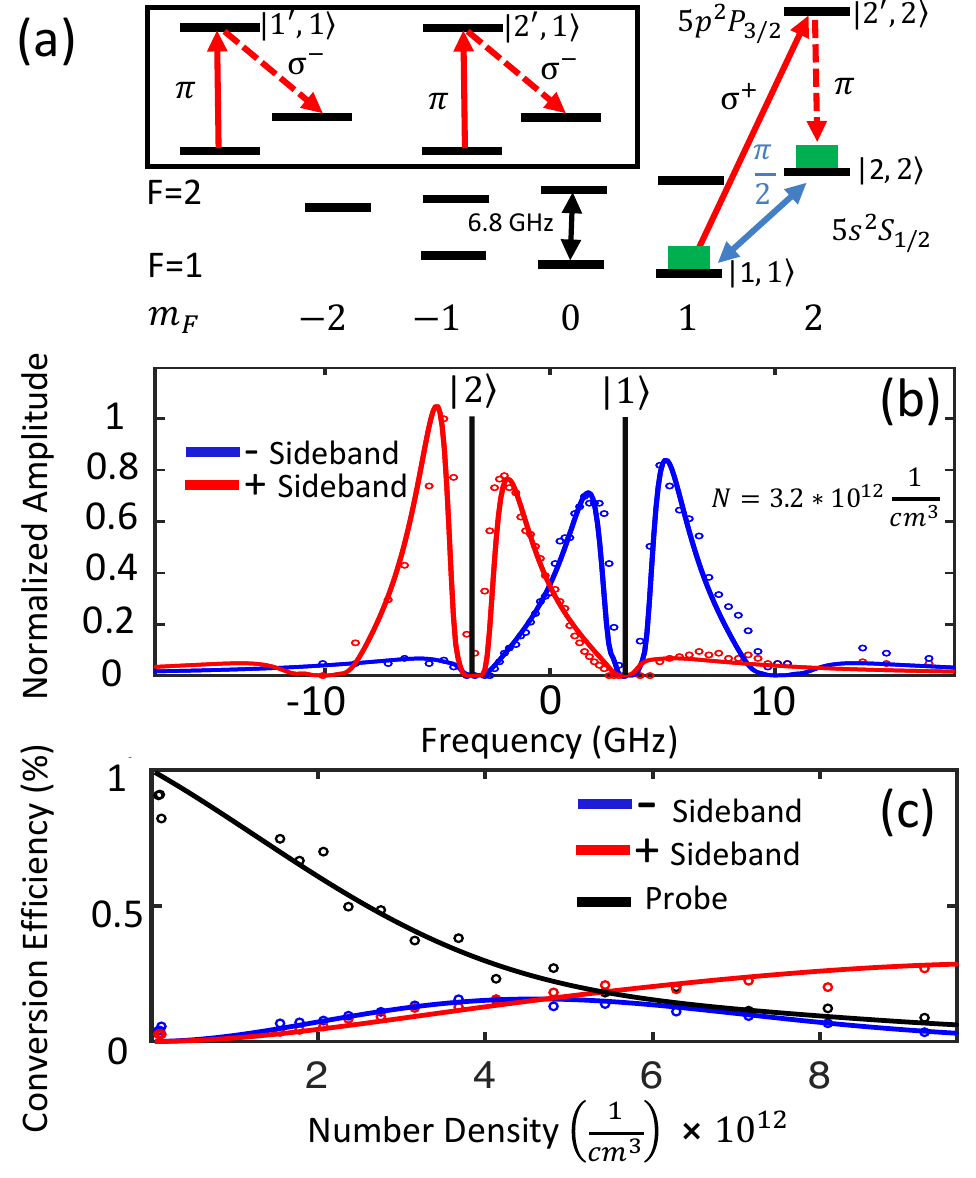}%
\end{center}
\vspace{-0.25in}
\caption{\label{fig:epsart}(a) Ground state manifold of $^{87}$Rb with transitions to an allowable excited state level. An example of a $\Lambda$ system needed for sideband generation is shown. (b) First order sideband amplitude as the probe frequency is swept across transitions from the $|2\rangle$ and $|1\rangle$ states. The open circles are experimental data, and the solid lines are from the numerical model, which is qualitatively fit to the experimental data from the FP etalon. (c) Probe and sideband amplitude as a function of rubidium density.}
\label{fig:concept}
\end{figure}

Using the setup shown in Fig. 2a, we send the light to a scanning FP etalon, and the amplitudes of the first order sidebands and probe light are measured. In Fig. 3b, we experimentally scan the frequency of the probe across the $|1\rangle$ and $|2\rangle$ resonances, measuring the amplitude of the sidebands, and results from our numerical model are plotted with the experimental data. In the model, we calculate the propagation of the sideband through the medium at the number density $3.2*10^{12}$ cm$^{-3}$ and determine the sideband amplitudes at the exit of the cell. Plots of sideband propagation as a function of distance are shown in the supplementary material. The amplitude of the sideband and probe light is severely reduced on resonance due to absorption by the atoms. Also of note, if the probe frequency is set outside of either resonance by the hyperfine splitting, one of the sidebands is again absorbed. For example, when the probe is tuned to about $-10.2$ GHz ($-6.8$ GHz detuned from the $|2\rangle$ resonance), the positive frequency sideband (red) is absorbed because it is generated at the frequency of the resonance. The negative frequency sideband (blue) is not absorbed because its frequency is far from resonance. However, its amplitude is reduced because the polarizability is inversely proportional to the optical detuning, as shown in the supplementary material. In Fig. 3c we measure the sideband conversion efficiency versus number density. The conversion efficiency is equal to $P_s/ P_c$, where $P_c$ is the power of the probe before it enters the vapor cell, and $P_s$ is the power of a particular sideband after leaving the cell. We set the probe frequency to be about halfway between the $|1\rangle$ and $|2\rangle$ resonances. The data and simulation of  Fig. 3b indicate that the two sidebands should be the same amplitude for this probe detuning. However, the optical pumping was not $100\%$ efficient in the experiment, and all the population was not initially in the $|2,2\rangle$ state. We found that to model this situation we could simply implement an imperfect $\pi/2$ pulse. Also, the model predicts the probe frequency was shifted $300$ MHz towards $|1\rangle$ indicating we were not exactly between the two resonances. With about $60\%$ of the atomic population remaining in the $|2,2\rangle$ state after the microwave pulse, the model shows good agreement with the data. With an imbalance in population, one sideband is absorbed more quickly than the other as it propagates through the cell, and the model clearly predicts this behavior.

Guided by our tabletop experiments and numerical simulations, we developed a compact gradiometer sensor package. A schematic of our sensor package is shown in Fig. 4a. The laser system is composed of two separate distributed feedback (DFB) lasers at $795$ nm wavelength and resonant with the D$1$ transition. The lasers are coupled to the sensor head by a polarization maintaining (PM) optical fiber. A high-speed ($<1 \ \mu$s) fiber-coupled electro-optical switch from Boston Applied Technologies (not shown) is used to switch the pump light on or off. The pump and probe beams are collimated to a $1$ cm $1/e^2$ diameter. The probe passes through a polarizing beam splitter ($\mathbf{PBS}$), and two cubic ($8\times8\times8$ mm$^{3}$ internal dimensions) $^{87}$Rb
vapor cells $\mathbf{A}$ and $\mathbf{B}$ separated by 4.4 cm. The pump is directed through the two cells in a direction orthogonal to the probe using a 50-50 non-polarizing beam splitter ($\mathbf{BS}$) and Mirror ($\mathbf{C})$.

In addition to  $^{87}$Rb, the vapor cells contain buffer gas with cell $\mathbf{A}$ nominally filled with $15$ Torr of $N_2$ pressure and cell $\mathbf{B}$ nominally filled with 30 Torr of $N_2$ pressure. Heaters made from a flexible printed circuit board (not shown) are glued to the unused wall of the vapor cells to control the rubidium vapor pressure. A linearly polarized microstrip antenna is positioned between the two cells to apply a microwave $\pi/2$ pulse at frequency $\Delta \nu$ in Eq.~1. The probe, after passing through the two cells, is retroreflected using a gold-coated mirror ($\mathbf{D}$) placed 2.2 cm away from the center cell $\mathbf{B}$.

\begin{figure}[t]
\begin{center}
\includegraphics[width=.5\textwidth,height=.40\textwidth]{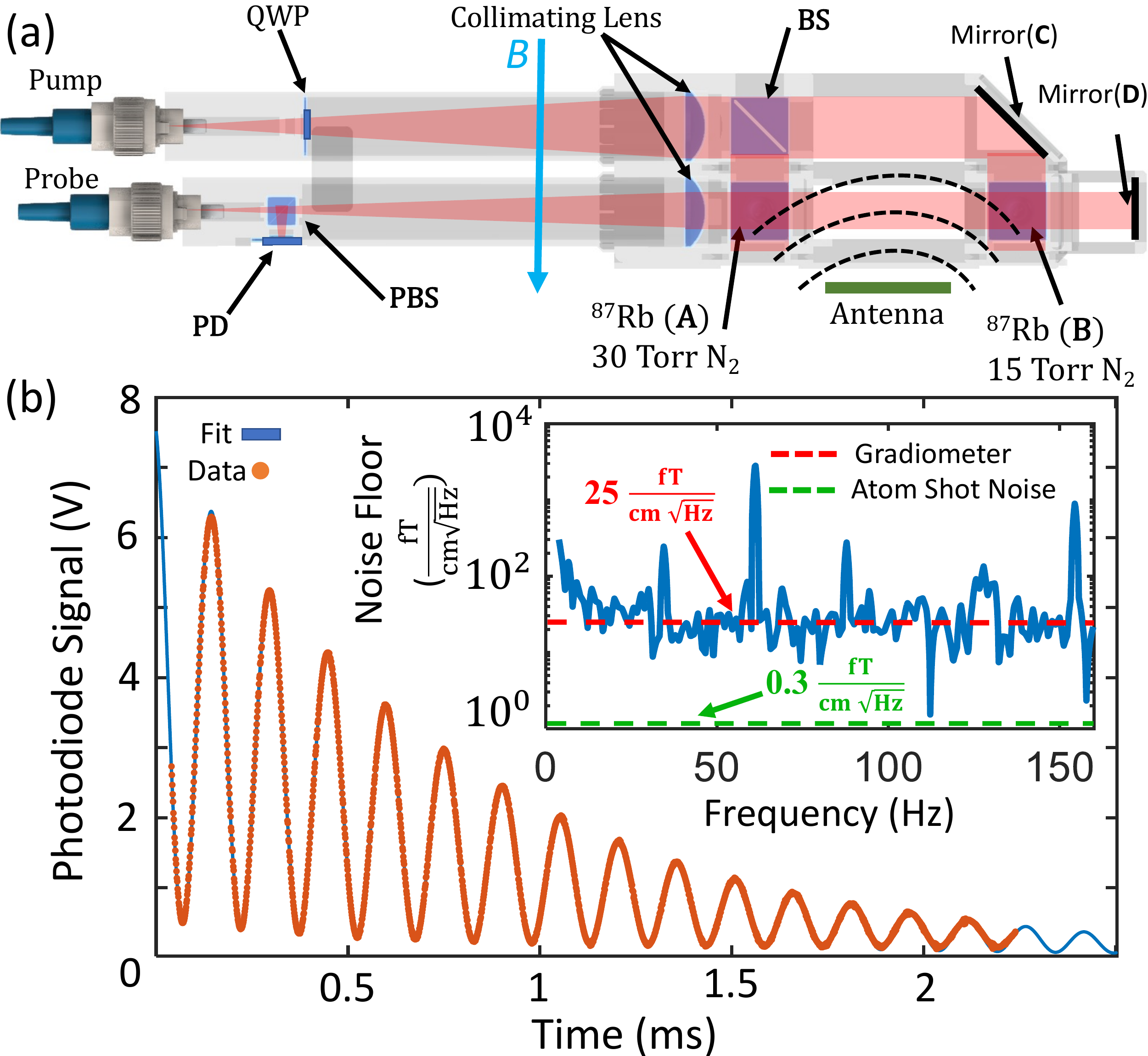}
\end{center}
\vspace{-0.25in}
\caption{\label{fig:epsart}(a) A schematic of the compact gradiometer sensor head. The probe makes two passes through vapor cells filled with $^{87}$ Rb vapor and nitrogen buffer gas. The sidebands  are separated from the probe by the $\mathbf{PBS}$ and measured on a $\mathbf{PD}$. (b) (Orange circles) $\mathbf{PD}$ output showing the beat-note observed after applying a $\pi/2$ pulse. (Blue) Numerical fit to experimental data with Eq.3. (Inset) Gradiometer noise floor obtained from the time series of beat-note frequencies in sequential cycles of gradiometer operation. Data was taken in an unshielded laboratory environment. }
\label{fig:concept}
\end{figure}

The distances between the cells ($\mathbf{A},\mathbf{ B}$) and the mirror ($\mathbf{D}$) were carefully chosen to ensure constructive interference of the relative phase between the probe and sidebands for forward and backward propagating probe light. Maximum sideband generation occurs at integer spacings of the wavelength of the microwave radiation. This discrete spacing relative to the mirror gives a 2.2-cm spacing between the mirror and Cell $\mathbf{A}$ and 6.6-cm spacing for the mirror to Cell $\mathbf{B}$, making the cell-to-cell separation (gradiometer baseline) 4.4 cm. After passing through the cells, the retroreflected light again passes through the $\mathbf{PBS}$ which directs the orthogonally polarized sideband light towards the $\mathbf{PD}$. 

The sensor is run in a pulsed mode with a 3 ms cycle time. During the first phase of the cycle (1.5 ms long),  the pump light is switched-on to spin polarize the atoms in both vapor cells. The pump laser wavelength is modulated at $200$ kHz to clear both the  $ 5s^{2} S_{1/2}~ |1\rangle$ and $|2\rangle$ ground state manifolds and transfer a majority of the atoms into the the $|2,2\rangle $ dark state. Due to limitations in the available pump optical power in the experiment ($4$ mW at the entrance of the sensor head), we estimate that roughly $60\%$  of the atoms were optically pumped into the $|2,2\rangle $ dark state in the two cells. After the optical pumping phase, the pump light is switched-off and a $\pi/2$ pulse is applied to generate sidebands. The sideband light is separated from the probe light by the $\mathbf{PBS}$, allowing only sideband light to pass to the $\mathbf{PD}$.

Interference between the sideband light from the two vapor cells generates a beat-note directly proportional to the gradient field between the two cells. Even in the absence of a significant gradient field, a beat-note with a $6$ kHz frequency is observed due to the nominal $15$ Torr differential in nitrogen buffer gas pressure between the two cells. The nitrogen buffer gas induced pressure shift for the ground state hyperfine transitions, $\Delta \nu_{BG}$,  is approximately $548$ Hz/Torr \cite{vanier1982}. This offset frequency allows the gradiometer to function in a near zero-gradient field environment.

Since the beat-note is produced by an interference of the electric fields from two exponentially decaying coherent light sources, the functional form of the photodiode output after the $\pi/2$ pulse is given by 

\begin{equation}
\resizebox{.9\hsize}{!}{$S(t) = {E_A^2}e^{{-\frac{2t}{T_{A}}}} + {E_B^2}e^{{-\frac{2t}{T_{B}}}} +2 E_AE_Be^{-t(\frac{1}{T_{A}}+\frac{1}{T_{B}})} \sin(2\pi f t+\phi)$}
\end{equation}
where $E_A$ and $E_B$ are the electric fields produced by the sidebands from cells $\mathbf{A}$ and $\mathbf{B}$ respectively, $T_{A}$ and $T_{B}$ are the hyperfine ground state relaxation rates in the two cells, and $f=\Delta \nu_{BG} + 3\gamma |\Delta \mathbf{B}|$ is the relative frequency difference between the two sidebands (beat-note frequency). Considering $\Delta \nu_{BG}$ to be a constant offset, the magnitude of the magnetic field gradient between the two cells is obtained from a measurement of $f$. The raw photodiode output was digitized and recorded using a 16-bit, 2 MSPS analog-to-digital converter (ADC). In post processing, the beat-note obtained in each cycle was fitted to the functional form $S(t)$ to extract the beat-note frequency. As seen in Fig. 4b, the fitted function closely overlaps with the experimental data. From the measurement of the beat-note frequency in each 3 ms cycle, a time series of the gradient field was obtained. From the Fourier transform of the gradient field time series, the gradiometer noise floor was calculated. Fig. 4b (inset) shows the gradiometer noise floor measured in the ambient laboratory background magnetic field to be  $25 \ fT/cm/\sqrt{Hz}$ . An estimate of the atom shot noise ($0.3 \ fT/cm/\sqrt{Hz}$) from Eq.~1, Ref\cite{allred2002high} is also shown. We expect the sensitivity to approach the atom shot noise limit by increasing the pump power to obtain unity spin polarization, improving the optics, and mitigating the technical noise. To improve the practical utility of our gradiometer, we also developed a second variant of the sensor package with
colinear pump and  probe beams that is described in the supplemental section.

In summary, we developed a novel method for direct optical observation of the magnetic field gradient between two spatially separated alkali vapor cells. The method relies on highly efficient sideband generation from the interaction between a linear probe beam and alkali atoms in a coherent superposition of magnetically sensitive hyperfine ground states. We developed a rigorous theoretical framework to accurately model our system and identify parameters to optimize its sensitivity. In addition, we demonstrate that our method is suitable for practical applications by developing an integrated, highly sensitive sensor package with readily available optical components. The sensor has the necessary characteristics to enable its adoption in a wide range of geophysical, scientific, and biomagnetic related applications. 

Author contributions. Conceptualization: P.S., Y.J., I.S., V.S.; Tabletop experiments: K.C., Y.W.; Modeling and Analysis: K.C., P.S., Y.W., Y.J.; Sensor design and construction: Y.W., V.S.; Theory: Y.J.; Paper writing: K.C., P.S., V.S. This work was funded by the Defense Advanced Research Projects Agency (DARPA) under the AMBIIENT program, contract No. 140D6318C0021. Approved for Public Release, Distribution Unlimited.  Sandia National Laboratories is a multimission laboratory managed and operated by National Technology and Engineering Solutions of Sandia, LLC, a wholly owned subsidiary of Honeywell International Inc., for the U.S. Department of Energy National Nuclear Security Administration under Contract No. DE-NA0003525. This paper describes objective technical results and analysis. Any subjective views or opinions that might be expressed in the paper do not necessarily represent the views of the U.S. Department of Energy or the U.S. Government.

\bibliography{apsCitationYJ}

\end{document}


\date{}
\maketitle
The cgs-unit derivation of Eq.~(2)  in the main text and the equations shown below can be found in Jau's unpublished manuscript \cite{Jau2007} as a separate supplemental PDF file. This original manuscript contains some English and compiling typos, but it provides the foundation of our modeling effort in this work. The undefined notations and mathematical formalisms in that manuscript are clarified by his coauthored book \cite{OPA} entitled ``Optically Pumped Atoms,'' Wiley-VCH (2010). 
For pedagogical purpose, we illustrate how we use Eq.~(2) in the main text to calculate the spatial evolution of the carrier and the optical sidebands. The optical carrier (probe laser beam) and sidebands can be represented as a frequency-quantized complex electric field $\mathbf{E(\zeta)}=\sum_\omega\widetilde{\mathbf{E}}_\omega(\zeta)|\omega\rangle e^{-i\omega t}$ and  $\widetilde{\mathbf{E}}(\zeta)=\sum_\omega\widetilde{\mathbf{E}}_\omega(\zeta)|\omega\rangle$, where $\omega$ is the optical frequency, $\widetilde{\mathbf{E}}_{\omega}(\zeta)$ is the position-dependent complex amplitude, $\zeta$ is the propagation distance, and $|\omega\rangle$ is a discrete frequency basis. For SI units, we can write the matrix elements of these operators using the following equations:
\begin{equation}\label{K}
\mathbf{K}|\omega\rangle = \frac{\omega}{c}\mathbf{1}|\omega\rangle,
\end{equation}
and
\begin{equation}
\resizebox{.55\hsize}{!}{$\label{chi}
\langle\omega'|\boldsymbol{\chi}|\omega\rangle=\frac{N}{\epsilon_0} \sum_{\mu\nu}\boldsymbol{\alpha}_{\mu\nu}(\omega)\tilde{\rho}_{\mu\nu}(\zeta)\delta_{(\omega'-\omega),\Omega_{\mu\nu}} e^{i\frac{(\omega'-\omega)}{c}\zeta}$}.
\end{equation}
Here, $c$ is the speed of light in vacuum, $\mathbf{1}$ is the unit dyadic operator, $\omega$ and $\omega'$ denote the initial and final frequency states of the optical electric field, $N$ is the alkali-metal atom number density, $\epsilon_0$ is the vacuum permitivity, $\boldsymbol{\alpha}_{\mu \nu}$ is the polarizability dyadic associated with two ground-state sublevels labeled by $\mu$ and $\nu$, $\tilde{\rho}_{\mu\nu}(\zeta)$ is a distance-dependent matrix element in multi-rotating frames of an atomic density-matrix $\rho$ in which  $\rho_{\mu\nu}=\tilde{\rho}_{\mu\nu}e^{-i\Omega_{\mu\nu}t}$, $\delta_{(\omega'-\omega),\Omega_{\mu\nu}}$ is the Kronecker delta, and $\Omega_{\mu\nu}$ is the angular frequency of an atomic coherence between the two ground-state sublevels. We find the polarizability operator to be described by
\begin{equation}\label{alpha}
\boldsymbol{\alpha}_{\mu \nu}(\omega)=\frac{1}{\hbar}\sum_{\bar{\mu}}  \frac{\mathbf{D}_{\mu\bar{\mu}}\mathbf{D}^\dag_{\bar{\mu}\nu} }{(\omega_{\bar{\mu}\nu}-\omega)-i\gamma_{\mu\nu}},
\end{equation}
where $\mathbf{D}_{\mu\bar{\mu}}$ is the element for electric dipole operator $\mathbf{D}$ defined by ground-state sublevel $\mu$ and excited-state sublevel $\bar{\mu}$. We can then numerically solve Eq.~(2) in the main text. 

Assuming we have a weak carrier beam incident on a warm vapor cell filled with $^{87}$Rb atoms, strong sidebands can be produced if the carrier (probe laser beam) is tuned to an appropriate detuning from the optical resonances. For illustration, we include only up to the second order sidebands in the model which means $\widetilde{\mathbf{E}}$ has five frequency components (The first and second order positive and negative sidebands and the carrier frequency). In this example, assuming the carrier light is $z$-polarized and moving along $y$, and the magnetic field is along $z$, we only need to consider the  $x$ and $z$ vector components of the electric field since there is no component of the electric field in the direction of propagation. From Eq.~(2) in the main text we define $\zeta=y$, and write out the components of the dyadics as $\textbf{K}=K_{xx}\mathbf{xx}+K_{zz}\mathbf{zz}$ and $\boldsymbol{\chi}=\chi_{xx}\mathbf{xx}+\chi_{zx}\mathbf{zx}+\chi_{xz}\mathbf{xz}+\chi_{zz}\mathbf{zz}$. Using Eq.~(1), we write
\begin{equation}
K_{xx}=K_{zz}=\frac{1}{c}  \begin{bmatrix}
\omega_{-2}&0&0&0&0\\
0 &\omega_{-1} & 0 & 0&0\\
0 & 0&\omega_0 & 0 &0\\
0  & 0 & 0&\omega_{1}&0 \\
0  & 0 & 0&0&\omega_{2}\end{bmatrix}
 \end{equation}
where  $(\omega_{-2},\omega_{-1},\omega_{0},\omega_{1},\omega_{2})=(\omega_0-2\Omega_{12},\omega_0-\Omega_{12},\omega_0,\omega_0+\Omega_{12},\omega_0+2\Omega_{12})$ and we recall that $\Omega_{\mu \nu}$ is the angular frequency of the atomic coherence between two ground-state hyperfine sublevels. In the experiment, the atoms are optically pumped to the end state so we use the two ground-state hyperfine sublevels $|1\rangle=|F=1,m_F=1\rangle$ and $|2\rangle=|F=2,m_F=2\rangle$ in our model. Using Eq.~(2) we write
 \begin{equation}
 \tiny
  \hspace{-0.80 in}
 \chi_{xx}= \frac{N}{\epsilon_0}
 \begin{bmatrix}
\alpha_{11}(\omega_{-2}) \tilde{\rho}_{11} + \alpha_{22}(\omega_{-2}) \tilde{\rho}_{22}&0&0&0&0 \\ 
0&\alpha_{11}(\omega_{-1}) \tilde{\rho}_{11} + \alpha_{22}(\omega_{-1}) \tilde{\rho}_{22} &0& 0&0\\ 
0&0 & \alpha_{11}(\omega_0) \tilde{\rho}_{11} + \alpha_{22}(\omega_0) \tilde{\rho}_{22} &0&0\\ 
0&0  &0& \alpha_{11}(\omega_{1}) \tilde{\rho}_{11} + \alpha_{22}(\omega_{1}) \tilde{\rho}_{22} &0\\
0&0&0&0&\alpha_{11}(\omega_{2}) \tilde{\rho}_{11} + \alpha_{22}(\omega_{2}) \tilde{\rho}_{22}
\end{bmatrix}
 \end{equation}

and
 
 \begin{equation} 
 \tiny
 \hspace{-0.4 in}
 \chi_{zx}= \frac{N}{\epsilon_0}
 \begin{bmatrix}
0& \alpha_{21}(\omega_{-1}) \tilde{\rho}_{21} e^{-i\Omega_{12}y}&0&0&0 \\ 
\alpha_{12}(\omega_{-2}) \tilde{\rho_{12}} e^{i\Omega_{12}y}&0 & \alpha_{21}(\omega_0) \tilde{\rho}_{21} e^{-i\Omega_{12}y} & 0&0\\ 
0&\alpha_{12}(\omega_{-1}) \tilde{\rho_{12}} e^{i\Omega_{12}y}  &0& \alpha_{21}(\omega_{1})\tilde{\rho_{21}} e^{-i\Omega_{12}y}&0\\ 
0&0  & \alpha_{12}(\omega_0) \tilde{\rho_{12}} e^{i\Omega_{12}y} & 0 &\alpha_{21}(\omega_{2}) \tilde{\rho}_{21} e^{-i\Omega_{12}y} \\
0&0&0&\alpha_{12}(\omega_{1}) \tilde{\rho_{12}} e^{i\Omega_{12}y}&0
\end{bmatrix}.
 \end{equation}
In a similar fashion, $\chi_{xz}$ and $\chi_{zz}$ can be constructed. In order to simplify the notation we let $\alpha_{\mu\mu}(\omega)=\alpha_{xx\mu\mu}(\omega)$ and $\alpha_{\mu\nu}(\omega)=\alpha_{zx\mu\nu}(\omega), \mu\neq\nu$, which can be calculated using Eq.~(3) . For SI-unit calculations one can use the dipole values and the Clebsch-Gordan Coefficients listed in Ref.~\cite{Steck} to construct $\mathbf{D}$ and calculate or use Eq.~(6-8) in Ref.~\cite{Jau2007} to calculate and then convert the unit from cm$^{3}$ to C $\cdot$ m$^{2} \cdot$ V$^{-1}$ . Also, $\alpha_{xx\mu\nu}(\omega)$ is the polarizability dyadic with optical transitions  $\sigma_+$ or $\sigma_-$ only due to selection rules for $x$ polarized light , and $\alpha_{zx\mu\nu}(\omega)$ is the case for a lambda system with $\pi$ transitions to $\sigma_+$ and $\sigma_-$  transitions. We can now expand  Eq.~(2) from the main text in matrix form. First we expand the dot product of the dyadics in the propagation equation as

  \begin{equation} 
 \hspace{-.8in}
 \mathbf{K}\cdot{(\mathbf{1}+\frac{\boldsymbol{\chi}}{2})}=
 \begin{bmatrix}
 K_{xx} &0&0\\ 
0&0&0\\ 
0&0 &K_{zz} \\ 
\end{bmatrix} 
+
 \begin{bmatrix}
 K_{xx}\frac{\chi_{xx}}{2}&0&K_{xx}\frac{\chi_{xz}}{2}\\ 
0& 0 &0\\ 
K_{zz} \frac{\chi_{zx} }{2}&0 &K_{zz} \frac{\chi_{zz}}{2} \\ 
\end{bmatrix} 
= 
 \begin{bmatrix}
  K_{xx} +K_{xx}\frac{\chi_{xx}}{2}&0&K_{xx} \frac{\chi_{xz}}{2}\\ 
0& 0&0\\ 
K_{zz}\frac{\chi_{zx} }{2}&0 &K_{zz}+K_{zz}\frac{ \chi_{zz}}{2} \\ 
\end{bmatrix} .
\end{equation}
Finally, we can write the propagation equation in matrix form as 

 \begin{equation} 
  \begin{bmatrix}
\frac{\partial{{\widetilde{E}_x}}}{\partial y}\\ 
\frac{\partial{{\widetilde{E}_y}}}{\partial y}\\ 
\frac{\partial{{\widetilde{E}_z}}}{\partial y} \\ 
\end{bmatrix}
=
i \begin{bmatrix}
  K_{xx} +K_{xx}\frac{\chi_{xx}}{2}&0&K_{xx} \frac{\chi_{xz}}{2}\\ 
0&0&0\\ 
K_{zz}\frac{\chi_{zx} }{2}&0 &K_{zz}+K_{zz}\frac{ \chi_{zz}}{2} \\ 
\end{bmatrix} 
\begin{bmatrix}
{\widetilde{E}_x}\\ 
{\widetilde{E}_y}\\ 
{\widetilde{E}_z}\\ 
\end{bmatrix}.
\end{equation}

 We then solve Eq.~(8)  numerically for the scenario that no significant sidebands are generated beyond 2nd-order. With modern computing power, the frequency basis can be easily extended for much higher-order sidebands if needed. The optical sidebands are generated in intervals of $\Omega_{12}$ as long as the atomic coherence $\tilde{\rho}_{12}$ and its complex conjugate $\tilde{\rho}_{21}$ are nonzero. In reality, the atomic density matrix $\rho$ is spatially dependent. However, if the optical pumping illumination is uniform across the vapor cell and the microwave radiation is perpendicular to the beam path, it is okay to assume $\rho$ to be the same along the entire beam path. If the microwave field has a non-zero projection along the beam path, a spatially dependent phase has to be implemented into $\rho$.
\begin{figure*}[t]
\vspace{-.5in}
\begin{center}
\includegraphics[width=.92\textwidth]{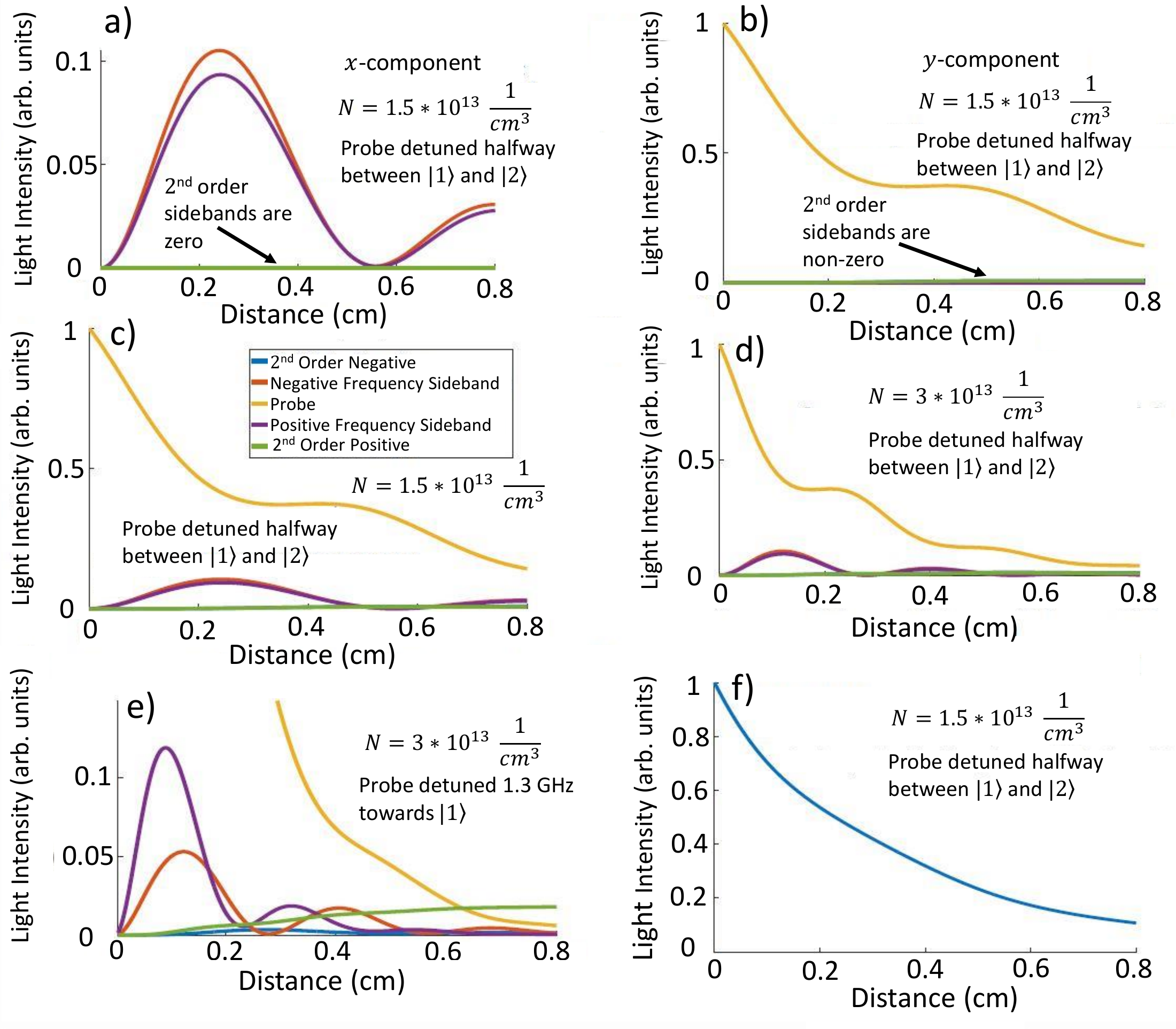}
\end{center}
\vspace{-.25in}
\caption{\label{fig:epsart}  Plots generated from the numerical model. The legend for the figure is shown in plot c. (a) $x$-component of the propagated light. (b) $y$-component of the propagated light. (c) $x$ and $y$ polarized components on the same plot. (d) Both $x$ and $y$ components for a higher number density. (e) Case where significant light is scattered into a second order sideband. (f) Total light Intensity (sideband and probe added up).}\end{figure*}

We use an $8^{th}$ order Runga-Kutta to solve Eq.~(8). Fig.~1 shows plots generated from solving the propagation equation for different starting parameters. The plots show probe (carrier) and sideband propagation through a vapor cell with a length of $0.8$ cm. At $z=0$, just as the light is entering the cell, the carrier beam has a total amplitude of 1. As the light propagates through the atoms, light is scattered into the first order sideband modes. From there the light can be further scattered into higher order sidebands or back to the carrier. Fig.~1(a) shows the $x$-polarized light as it propagates through the medium. Notice that only the first order sidebands are scattered into the $x$-direction since the incoming carrier light is $z$-polarized. The carrier light is frequency detuned midway between the $F=1$ and $F=2$ levels.
Fig.~1(b) shows the light polarized in the $z$-direction, which is the carrier and second order sidebands. This makes sense, since the sideband generation process scatters the light into an orthogonal mode, and light scattered from the first order sidebands will be scattered into a mode that is parallel to the incoming carrier polarization.  Fig.~1(c) shows all of the light on the same plot.  Fig.~1(d) is the same as Fig.~1(c) but with a higher number density. In this case the optical density is larger and the light is absorbed more quickly as it propagates. In Fig.~1(e) the probe is shifted $1.3$ GHz towards the $F=1$ level. This position is interesting because by the time the light makes it through the cell, most of the light is actually in one of the second order sidebands (green). Also notice how the green sideband doesn't decay over the length of the cell as quickly as the other sidebands due to the fact that it is farther away from the resonances.  Fig.~1(f) shows the total light intensity, added up, as it passes through the cell. The changes in slope of the plot are due to the fact that scattered light is a different distance away from resonance. 
\bigskip

\centerline{\fontsize{22}{15}\selectfont{Colinear Pump-Probe Variant}}
\fontsize{11}{15}\selectfont
\hfill

To improve the practical utility of our gradiometer, we developed a second variant of the sensor package with colinear pump and probe beams to minimize the sensor footprint and reduce its dead zones \cite{kish2010}. A dead zone is defined as the orientation of the sensor package with respect to the background field direction in which the sensor is inoperable. The orthogonal pump-probe variant enters a dead zone when the background field is substantially perpendicular to the direction of the pump beam through the vapor cell. The dead zone occurs because the atoms are no longer efficiently pumped into the $|2,2\rangle$ state. In contrast, the modified colinear pump-probe variant reduces the planar dead zone to a single axial dead zone. Additional strategies to remove the single axis dead zone were found to be feasible and we plan to describe the results in a future paper and pending patent.

\begin{figure}[h]
\begin{center}
\vspace*{-.3 cm}
\includegraphics[width=1\textwidth]{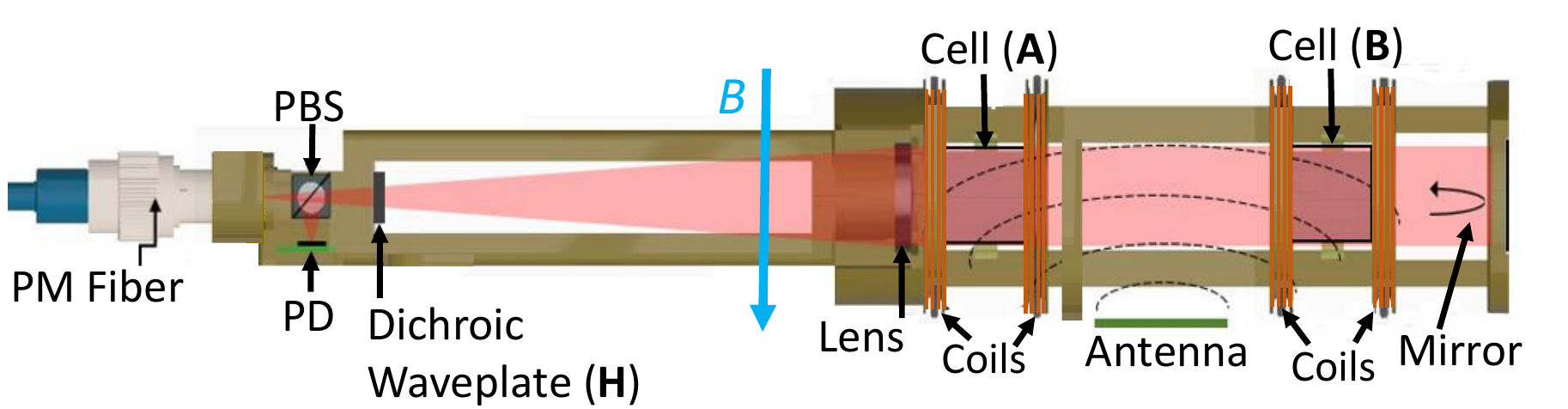}
\end{center}
\vspace{-.25 in}
\caption{\label{fig:epsart} A schematic of the gradiometer sensor head in the collinear pump/probe configuration.}
\label{fig:concept}
\end{figure}

In the colinear pump-probe variant (Fig.~5), the 795 nm pump and 780 nm probe beams were combined into a single polarization maintaining fiber before entering the sensor package. Both beams pass through a polarizing beam splitter ($\mathbf{PBS}$) and subsequently through a dichroic waveplate $\mathbf{H}$ ($\lambda/4 \ 795$ nm, $\lambda/2 \ 780$ nm) which circularly polarizes the $795$ nm pump light while leaving the $780$ nm probe light linearly polarized \cite{Johnson2010}. Vapor cells $\mathbf{A}$ and $\mathbf{B}$ are each surrounded by a pair of Helmholtz coils which provide a $300 \ \mu$T bias field ($5-10$ times Earth's field) during the optical pumping phase of the cycle. After pumping, the bias field is adiabatically switched-off to align the spins with the background field before applying the microwave $\pi/2$ pulse. Aside from these changes the sensor operation between the two variants remains unchanged. No noticeable change in amplitude of the beat note signal was observed in the colinear variant. 

\bibliography{apsCitationYJ}